\documentclass[10pt,journal]{IEEEtran}
\usepackage{amsmath}
\usepackage{amssymb}
\usepackage{amsfonts}
\usepackage{graphicx}
\usepackage{float}

\input{tcilatex}

\begin{document}

\title{Frequency-Domain Analysis of Nonlinear and Linear Integrators}
\author{Xinhua Wang\thanks{%
Xinhua Wang is with the Department of Mechanical and Aerospace Engineering,
Monash University, Clayton, VIC 3800, Australia (e-mail:
wangxinhua04@gmail.com).}}
\maketitle

\begin{abstract}
In this paper, frequency-domain analysis based on frequency sweep method is
presented for a nonlinear double integrator and a new linear integrator. All
the two types of integrators can estimate the onefold and double integrals
of a signal synchronously. With respect to the linear double integrator, the
nonlinear integrator has better estimation performance and stronger
robustness. Importantly, the integrator parameters can be regulated from the
frequency-domain analysis.
\end{abstract}


\section{Introduction}

\setcounter{page}{1}This paper focuses on the frequency-domain analysis of
two types of integrators, which can estimate the onefold and double
integrals of a signal synchronously.

Integrals are important components in almost all engineering applications.
The problem of double integral is that of estimating the number $I_{2}\left(
a\right) =\int_{0}^{t}\int_{0}^{s}a\left( \sigma \right) d\sigma d\tau $
with $\left[ 0,t\right] $ a finite time interval. Obtaining the double
integral of a signal is crucial for many kinds of systems, especially for
Inertial Navigation System (INS).

The usual observers or differentiators [1]-[4] can estimate the derivatives
of the signal, but not its multiple integrals. There are several linear
approximated methods to estimate onefold integral [5-8]: Romberg
integration, Gaussian quadrature, extended Simpson's rule, fractional-order
integrator. In all of the aforementioned studies, there is no stability
analysis. Furthermore, they are easily disturbed by stochastic noise
(especially non-white noise), and the drift phenomena occur in such systems.
In [9], a fractional-order integrator is proposed to approximate the
irrational fractional-order integrator $1/s^{m}$. However, the condition of $%
0<m<1$ limits the application of the fractional-order integrator. Recently
years, Kalman filter is used to handle the separation of probabilistic noise
and to estimate the position and velocity from the acceleration measurement
[10]-[12]. However, for Kalman filter, it is assumed that the process noise
covariance and measurement noise covariance are zero-mean Gaussian
distributed, and the process noise covariance is uncorrelated to the
estimation error. These assumptions are different from the real noise in
signal. The inaccurate noise information in sensed accelerations may lead to
the estimate drifts of position and velocity.

In [13], a nonlinear double integrator was presented based on finite-time
stability [14, 15]. The proposed double integrator can estimate the onefold
and double integrals of a signal synchronously, and the stability and
robustness in time domain were analyzed. The merits of the presented double
integrator include its finite-time stability, ease of parameters selection,
sufficient stochastic noise rejection and almost no drift phenomenon. The
theoretical results are confirmed by an experiment on a quadrotor aircraft
to estimate the position and velocity from the acceleration measurement. The
nonlinear double integrator leads to perform rejection of low-level
persistent disturbances. However, no robustness for high-frequency noise is
analyzed, i.e., no frequency-domain analysis on the effect of high-frequency
noise is considered. In fact, stochastic high-frequency noise exist in
almost all signals. Therefore, with respect to the ability of rejecting of
low-level persistent disturbances, the strong robustness of reducing
high-frequency noise is also necessary. And the parameters selection based
on frequency-domain analysis is required.

In this paper, based on singular perturbation technique [16, 17], a linear
double integrator is presented to estimate the onefold and double integrals
of a signal synchronously. Moreover, using frequency sweep method,
frequency-domain analysis is presented for the nonlinear double integrator
in [13] and the linear integrator. From the frequency-domain analysis,
comparing to the linear double integrator, the nonlinear integrator has the
better estimation performance and stronger robustness. Also the observer
parameters are more easily to be chosen from the frequency-domain analysis
with respect to the analysis in time domain.

\section{Nonlinear and linear double integrators}

\subsection{Nonlinear double integrator}

A nonlinear double integrator has been presented in [13], i.e., for system

\begin{eqnarray}
\dot{x}_{1} &=&x_{2};\dot{x}_{2}=x_{3};  \notag \\
\varepsilon ^{4}\dot{x}_{3} &=&-\sum\limits_{{i=1}}^{{2}}k_{i}\left\vert
\varepsilon ^{i}x_{i}\right\vert ^{\alpha _{i}}sign\left( x_{i}\right)
\notag \\
&&-k_{3}\left\vert x_{3}-a\left( t\right) \right\vert ^{\alpha
_{3}}sign\left( x_{3}-a\left( t\right) \right)
\end{eqnarray}%
if input signal $a\left( t\right) $ is the continuous and first-order
derivable, then there exist $\gamma >1$ and $\Gamma >0$, such that, for $%
t\geq \varepsilon \Gamma \left( \Xi (\varepsilon )e\left( {0}\right) \right)
$,

\begin{equation}
\left\vert x_{i}-a_{i}\left( t\right) \right\vert \leq L\varepsilon ^{\alpha
_{1}\gamma -i},i=1,2,3
\end{equation}%
where $a_{1}\left( t\right) =\int_{0}^{t}\int_{0}^{\sigma _{2}}a\left(
\sigma _{1}\right) d\sigma _{1}d\sigma _{2}$, $a_{2}\left( t\right)
=\int_{0}^{t}a\left( \sigma _{1}\right) d\sigma _{1}$; $x_{1}\left( 0\right)
=a_{1}\left( 0\right) $, $x_{2}\left( 0\right) =a_{2}\left( 0\right) $, $%
x_{3}\left( 0\right) =a_{3}\left( 0\right) $; $\varepsilon \in \left(
0,1\right) $ is the perturbation parameter; $L$ is some positive constant; $%
\gamma =(1-\beta )/\beta >1$, and $\beta \in \left( 0,\alpha _{1}/(\alpha
_{1}+4)\right) $; $\alpha _{1},\alpha _{2},\alpha _{3}$ satisfy:

\begin{equation}
\alpha _{3}\in (0,1),\alpha _{2}=\frac{\alpha _{3}}{2-\alpha _{3}},\alpha
_{1}=\frac{\alpha _{3}}{3-2\alpha _{3}}
\end{equation}%
$k_{1},k_{2},k_{3}>0$ are selected such that

\begin{equation}
k_{1}>0,k_{3}>0,k_{2}>\varepsilon ^{3\alpha _{3}}\frac{k_{1}}{k_{3}}
\end{equation}%
$e_{i}=x_{i}-a_{i}\left( t\right) $, $i=1,2,3$; $e=[%
\begin{array}{ccc}
e_{1} & e_{2} & e_{3}%
\end{array}%
]^{{T}}$; $\Xi (\varepsilon )=diag\{\varepsilon ,\varepsilon
^{2},\varepsilon ^{3}\}$.

In nonlinear double integrator (1), $x_{3}$ tracks the input signal $a(t)$, $%
x_{2}$ and $x_{1}$ estimate the onefold and double integrals of signal $a(t)$%
, respectively. From Theorem 1 in [13], nonlinear double integrator (1)
leads to perform rejection of low-level persistent disturbances. However, no
frequency-domain analysis on the effect of high-frequency noise is
considered. In fact, high-frequency noise exist in almost all signals.
Therefore, the frequency-domain analysis for the nonlinear double integrator
is inevitable.

As we know, a linear system is easy to perform frequency-domain analysis
with respect to nonlinear one. In the following, based on the design of
nonlinear double integrator, a simple linear double integrator will be
designed (when $\alpha _{3}=1$), and Theorem 1 is presented as follow.

\subsection{Linear double integrator}

\emph{Theorem 1:} For system

\begin{eqnarray}
\dot{x}_{1} &=&x_{2};\dot{x}_{2}=x_{3};  \notag \\
\varepsilon ^{4}\dot{x}_{3} &=&-k_{1}\varepsilon x_{1}-k_{2}\varepsilon
^{2}x_{2}-k_{3}\left( x_{3}-a\left( t\right) \right)
\end{eqnarray}%
if input signal $a\left( t\right) $\ is continuous, integrable and
first-order derivable, then

\begin{equation}
\underset{\varepsilon \rightarrow 0}{\lim }x_{i}=a_{i}\left( t\right)
,i=1,2,3
\end{equation}%
where $a_{1}\left( t\right) =\int_{0}^{t}\int_{0}^{\tau }a\left( s\right)
dsd\tau $, $a_{2}\left( t\right) =\int_{0}^{t}a\left( \tau \right) $, $%
a_{3}\left( t\right) =a\left( t\right) $; $x_{i}\left( 0\right) =a_{i}\left(
0\right) $, $i=1,2$; $\varepsilon \in \left( 0,1\right) $ is the
perturbation parameter; $k_{1},k_{2},k_{3}>0$ are selected such that

\begin{equation}
k_{1}>0,k_{3}>0,k_{2}>\varepsilon ^{3}\frac{k_{1}}{k_{3}}
\end{equation}

In linear double integrator (5), $x_{3}$ tracks the input signal $a(t)$, $%
x_{2}$ and $x_{1}$ estimate the onefold and double integrals of signal $a(t)$%
, respectively.

\emph{Proof of Theorem 1:}\textbf{\ }The Laplace transformation of Eq. (5)
can be obtained as follow:

\begin{eqnarray}
sX_{1}\left( s\right) &=&X_{2}\left( s\right) ;sX_{2}\left( s\right)
=X_{3}\left( s\right) ;  \notag \\
s\varepsilon ^{4}X_{3}\left( s\right) &=&-\sum\limits_{{i=1}}^{{2}%
}k_{i}\varepsilon ^{i}X_{i}\left( s\right)  \notag \\
&&-k_{3}\left( X_{3}\left( s\right) -A\left( s\right) \right)
\end{eqnarray}%
where $X_{i}\left( s\right) $ and $A\left( s\right) $ denote the Laplace
transformations of $x_{i}$ and $a\left( t\right) $, respectively, and $s$
denotes Laplace operator. From (8), we obtain

\begin{equation}
X_{i}\left( s\right) =\frac{X_{j}\left( s\right) }{s^{j-i}},i=1,2,3,j\in
\left\{ 1,2,3\right\}
\end{equation}

Therefore, Eq. (8) can be written as

\begin{eqnarray}
s^{3-j+1}\varepsilon ^{4}X_{j}\left( s\right) &=&-\sum\limits_{{i=1}}^{{2}%
}k_{i}\varepsilon ^{i}\frac{X_{j}\left( s\right) }{s^{j-i}}  \notag \\
&&-k_{3}\left( \frac{X_{j}\left( s\right) }{s^{j-3}}-A\left( s\right) \right)
\end{eqnarray}

Then, it follows that

\begin{equation}
\frac{X_{j}\left( s\right) }{A\left( s\right) }=\frac{k_{3}}{%
s^{3-j+1}\varepsilon ^{4}+\sum\limits_{{i=1}}^{{2}}\frac{k_{i}\varepsilon
^{i}}{s^{j-i}}+\frac{k_{3}}{s^{j-3}}}
\end{equation}%
i.e.,

\begin{equation}
\frac{X_{j}\left( s\right) }{A\left( s\right) }=\frac{s^{j-1}k_{3}}{%
s^{3}\varepsilon ^{4}+\sum\limits_{{i=1}}^{{2}}s^{i-1}k_{i}\varepsilon
^{i}+s^{2}k_{3}}
\end{equation}

Therefore, we obtain

\begin{equation}
\underset{\varepsilon \rightarrow 0}{\lim }\frac{X_{j}\left( s\right) }{%
A\left( s\right) }=s^{j-3}
\end{equation}%
where $j\in \left\{ 1,2,3\right\} $. It means that $x_{i}$ approximates $%
a_{i}\left( t\right) $ for $1\leq i\leq 3$.

Furthermore, the denominator of Equation (12) is required to be Hurwitz,
i.e., polynomial $s^{3}+\frac{k_{3}/\varepsilon ^{3}}{\varepsilon }s^{2}+%
\frac{k_{2}}{\varepsilon ^{2}}s+\frac{k_{1}}{\varepsilon ^{3}}$ is Hurwitz.
It is equivalent that $s^{3}+\frac{k_{3}}{\varepsilon ^{3}}%
s^{2}+k_{2}s+k_{1} $ should be Hurwitz. For arbitrary $\varepsilon \in
\left( 0,1\right) $, from the Routh-Hurwitz Stability Criterion, polynomial $%
s^{3}+\frac{k_{3}}{\varepsilon ^{3}}s^{2}+k_{2}s+k_{1}$ is Hurwitz if $%
k_{1}>0,k_{3}>0,k_{2}>\varepsilon ^{3}k_{1}/k_{3}$. This concludes the
proof. $\blacksquare $

From Theorem 1, the presented linear double integrator can't guarantee to
perform rejection of high-frequency noise. In fact, the robustness for the
effect of high-frequency noise exist. With respect to nonlinear double
integrator (1), the advantage of linear double integrator (5) is its simple
implementation.

In the next section, the frequency-domain analysis will be presented for the
nonlinear and linear double integrators.

\section{Frequency-domain analysis based on frequency sweep}

In a practical problem, high-frequency noises exist in measurement signal $%
a(t)$. The following analysis concerns the robustness behaviors of the
nonlinear and linear double integrators under high-frequency noises.

For the nonlinear double integrator, an extended version of the frequency
response method, frequency-sweep method [18, 19], can be used to
approximately analyze and predict the nonlinear behaviors of the nonlinear
integrator. Even though it is only an approximation method, the desirable
properties it inherits from the frequency response method, and the shortage
of other, systematic tools for nonlinear integrator analysis, make it an
indispensable component of the bag of tools of practicing control engineers.
By frequency-sweep method, we can find that the nonlinear double integrator
leads to perform precise estimation of integrals and strong rejection of
high-frequency noise.

The test of frequency characteristic can be implemented by Bode plot
fitting. For linear or nonlinear double integrators (1) or (5), let the
input signal be

\begin{equation}
a(t)=A_{m}\sin (\omega t)
\end{equation}%
where $A_{m}$ and $\omega $ are the amplitude and angular rate of the input
signal, respectively. Suppose the output of the double integrator can be
expressed as

\begin{eqnarray}
y(t) &=&A_{f}\sin (\omega t+\varphi )  \notag \\
&=&\left[
\begin{array}{cc}
\sin (\omega t) & \cos (\omega t)%
\end{array}%
\right] \left[
\begin{array}{c}
A_{f}\cos \varphi \\
A_{f}\sin \varphi%
\end{array}%
\right]
\end{eqnarray}%
where $A_{f}$, $\omega $ and $\varphi $ are the amplitude, angular rate and
phase\ of the output signal, respectively. Let $t=0$, $h$, $2h$, $\cdots $, $%
nh$, and

\begin{eqnarray}
Y^{T} &=&\left[
\begin{array}{cccc}
y(0) & y(h) & \cdots & y(nh)%
\end{array}%
\right]  \notag \\
\Psi ^{T} &=&\left[
\begin{array}{cccc}
\sin (\omega 0) & \sin (\omega h) & \cdots & \sin (\omega nh) \\
\cos (\omega 0) & \cos (\omega h) & \cdots & \cos (\omega nh)%
\end{array}%
\right]  \notag \\
c_{1} &=&A_{f}\cos \varphi ,c_{2}=A_{f}\sin \varphi
\end{eqnarray}%
where $h$ is the step size. From (15) and (16), it follows that

\begin{equation}
Y=\Psi \left[
\begin{array}{cc}
c_{1} & c_{2}%
\end{array}%
\right] ^{T}
\end{equation}

From the least square method, $c_{1}$ and $c_{2}$ can be obtained as follow:

\begin{equation}
\left[
\begin{array}{cc}
\hat{c}_{1} & \hat{c}_{2}%
\end{array}%
\right] ^{T}=(\Psi ^{T}\Psi )^{-1}\Psi ^{T}Y
\end{equation}

For the angular rate $\omega $, the amplitude and phase of the output signal
are, respectively, written as

\begin{equation}
\hat{A}_{f}=\sqrt{\hat{c}_{1}^{2}+\hat{c}_{2}^{2}},\hat{\varphi}=\arctan (%
\hat{c}_{2}/\hat{c}_{1})
\end{equation}

Therefore, the amplitude frequency characteristic can be described as

\begin{equation}
\hat{M}=20\lg (\hat{A}_{f}/A_{m})=20\lg (\sqrt{\hat{c}_{1}^{2}+\hat{c}%
_{2}^{2}}/A_{m})
\end{equation}%
and the phase frequency characteristic is the phase error between the output
and input, and it can be described as

\begin{equation}
\varphi _{e}=\varphi _{out}-\varphi _{in}=\arctan (\hat{c}_{2}/\hat{c}_{1})
\end{equation}

The angular frequency sequence $\{\omega _{k}\}$, where $k=1,\cdots ,n$, is
selected in the interested frequency bandwidth. For each angular frequency
the frequency bandwidth, the above frequency-sweep method is adopted to
obtain the values of amplitude and phase, respectively. Accordingly, the
Bode plots of the frequency-domain characteristic can be described.

In this frequency-domain analysis, $\omega =2\pi f$, where $f=0.1:0.5:100$; $%
h=0.001$; $k=1:1:50000$. Then input signal is $a(k)=A_{m}\sin (2\pi fkh)$.

In the following, using frequency-sweep method, we will analyze the effects
of the observer parameters on the estimate performances and robustness.

\subsection{Frequency characteristics with different $\protect\varepsilon $\
and $\protect\alpha _{3}$}

For the nonlinear and linear double integrators (1) and (5), the parameters
are selected as follows: $k_{1}=0.1$, $k_{2}=0.1$, $k_{3}=1$; $A_{m}=1$; $%
\alpha _{3}=\alpha =0.3,0.5,1$, respectively; $R=1/\varepsilon =3,4,5$,
respectively. The Bode plots of the frequency-domain characteristics with
different $\varepsilon $\ and $\alpha _{3}=\alpha $ are described in Figs.
1(a), 1(b) and 1(c), respectively: Fig. 1(a) presents the frequency
characteristic of signal tracking; Figs. 1(b) and 1(c) present the frequency
characteristics of onefold and double integral estimations, respectively. We
can find that high-frequency noise can be reduced sufficiently. Moreover,
decreasing parameter $\varepsilon $, the cut-off frequency become larger.
The smaller $\varepsilon $ is, the signal in wider frequency bandwidth can
be estimated. However, much noise will pass through the observer. On the
other hand, increasing parameter $\varepsilon $, the cut-off frequency
become smaller, much noise will be rejected.

Importantly, from Figs. 1(a), 1(b) and 1(c), changing parameter $\alpha
_{3}\in (0,1]$, the amplitude frequency characteristics almost don't be
affected, but the phase frequency characteristics are affected obviously:
when $\alpha _{3}$ approaches to $1$, the phase frequency characteristic
curves decay slowly near the cut-off frequency, and phase delay and
chattering exist. On the other hand, decreasing parameter $\alpha _{3}\in
(0,1)$, the phase frequency characteristic curves decay rapidly at the
cut-off frequency. Relatively smaller $\alpha _{3}\in (0,1)$ can obtain more
precise estimations and stronger robustness.

\subsection{Frequency characteristics with the change of $A_{m}$}

For nonlinear double integrator (1), the parameters are selected as follows:
$\alpha _{3}=0.3$; $R=1/\varepsilon =3$; $k_{1}=0.1$, $k_{2}=0.1$, $k_{3}=1$%
; $A_{m}=5,1,0.5$, respectively. The Bode plots of the frequency-domain
characteristics with the change of $A_{m}$ are described in Figs. 2(a), 2(b)
and 2(c), respectively: Fig. 2(a) presents the frequency characteristic of
signal tracking; Figs. 2(b) and 2(c) present the frequency characteristics
of onefold and double integral estimations, respectively. It is found that,
when the magnitude of input signal $A_{m}$ is larger, the cut-off frequency
is relatively smaller, and much noise is reduced sufficiently; when the
magnitude $A_{m}$ is smaller, the cut-off frequency is relatively larger,
and the signal in wider frequency bandwidth can be estimated.

\bigskip

\emph{Remark 1:} Comparing with ideal integral operators $1/s$ and $1/s^{2}$%
, not only the double integrators can obtain their estimations precisely,
but also the high-frequency noise is reduced sufficiently. Parameter $%
\varepsilon $ affects the low-pass frequency bandwidth: Decreasing the
perturbation parameter $\varepsilon $, the low-pass frequency bandwidth is
larger, the estimation precision becomes better, and relatively higher
frequency noise can be reduced; on the other hand, increasing perturbation
parameter $\varepsilon $, the low-pass frequency bandwidth is smaller, much
noise can be reduced sufficiently (See the cases of $R=1/\varepsilon =3,4,5$
in Figure 1, respectively). Parameter $\alpha _{3}\in (0,1]$ affects the
decay speed of frequency characteristic curves near the cut-off frequency
(See the cases of $\alpha _{3}=\alpha =0.3,0.5,1$ in Figure 1,
respectively): Smaller $\alpha _{3}\in (0,1]$ can obtain more precise
estimations; Larger $\alpha _{3}\in (0,1]$ can restrain much noise, however,
estimation delay phenomena exist.

\section{Computational analysis and simulations}

In this section, simulation results are presented in order to observe the
performances of the nonlinear and linear double integrators. We consider the
simulations of the following case: Double integrators for a input signal
with high-frequency noise. In the simulations, the function of $-0.1\times
3.14^{2}\times \sin (3.14t)$ is selected as reference signal $a_{03}(t)$.
Therefore, $a_{02}=\int_{0}^{t}a_{03}(\sigma )d\sigma =0.1\times 3.14\times
\cos (3.14t)$, and $a_{01}=\int_{0}^{t}\int_{0}^{s}a_{03}\left( \sigma
\right) d\sigma d\tau =0.1\sin (3.14t)$.

Here, the following high-frequency noise $\delta (t)$\ is selected (See the
noise in Fig. 3(a)): $0.1\sin (10t)+0.1\cos (10t)+0.05\sin (50t)+0.05\cos
(50t)$.

Therefore, the input signal $a(t)=a_{03}(t)+\delta (t)$.\ From the
frequency-domain analysis, the observer parameters can be selected as
follows: $R=1/\varepsilon =5$, $k_{1}=0.1$, $k_{2}=0.1$, $k_{3}=1$, $\alpha
_{3}=\alpha =0.3$, $\alpha _{2}=\frac{\alpha }{2-\alpha },\alpha _{1}=\frac{%
\alpha }{3-2\alpha }$; the initial value of the observer is ($x_{1}\left(
0\right) =0,x_{2}\left( 0\right) =1,x_{3}\left( 0\right) =0$). In the double
integrators (1) or (5), $x_{3}$ tracks signal $a_{03}(t)$, $x_{2}$ and $%
x_{1} $ estimate the onefold and double integrals of signal $a_{03}(t)$,
respectively.

Signal $a_{03}(t)$ tracking, the onefold and double integral estimations in
20 seconds are presented in Fig. 3. Fig. 3(a) provides signal $a_{03}\left(
t\right) $ with stochastic noise. Fig. 3(b) describes $a_{03}(t)$ tracking.
Figs. 3(c) and 3(d) present the onefold and double integral estimations,
respectively. Figs. 4(a)-4(d) describe the onefold and double integral
estimations in 2000 seconds.

From the above simulations, despite the existence of the intensive
high-frequency noise, the nonlinear double integrator showed the promising
estimation ability and robustness. Furthermore, from Figs. 4(a)-4(d), no
drift phenomenon happened in the long-time estimations.

Figs. 5 and 6 describe the estimations of onefold and double integrals by
the linear double integrator (5) (when $\alpha _{3}=1$), in 20s and 2000s,
respectively. From Figs. 5 and 6, the obvious estimation delay and slow
convergence of double integral exist.

\section{Conclusions}

Based on frequency sweep method, frequency-domain analysis is proposed for
the nonlinear double integrator and the linear integrator. All the two types
of integrators have the strong robustness for the effect of high-frequency
noise. Comparing to the linear double integrator, the nonlinear integrator
has the better estimation performance and stronger robustness. Importantly,
the integrator parameters are easily to be chosen from the frequency-domain
analysis with respect to the analysis in time domain.

\begin{figure}[H]
\begin{center}
\includegraphics[width=3.60in]{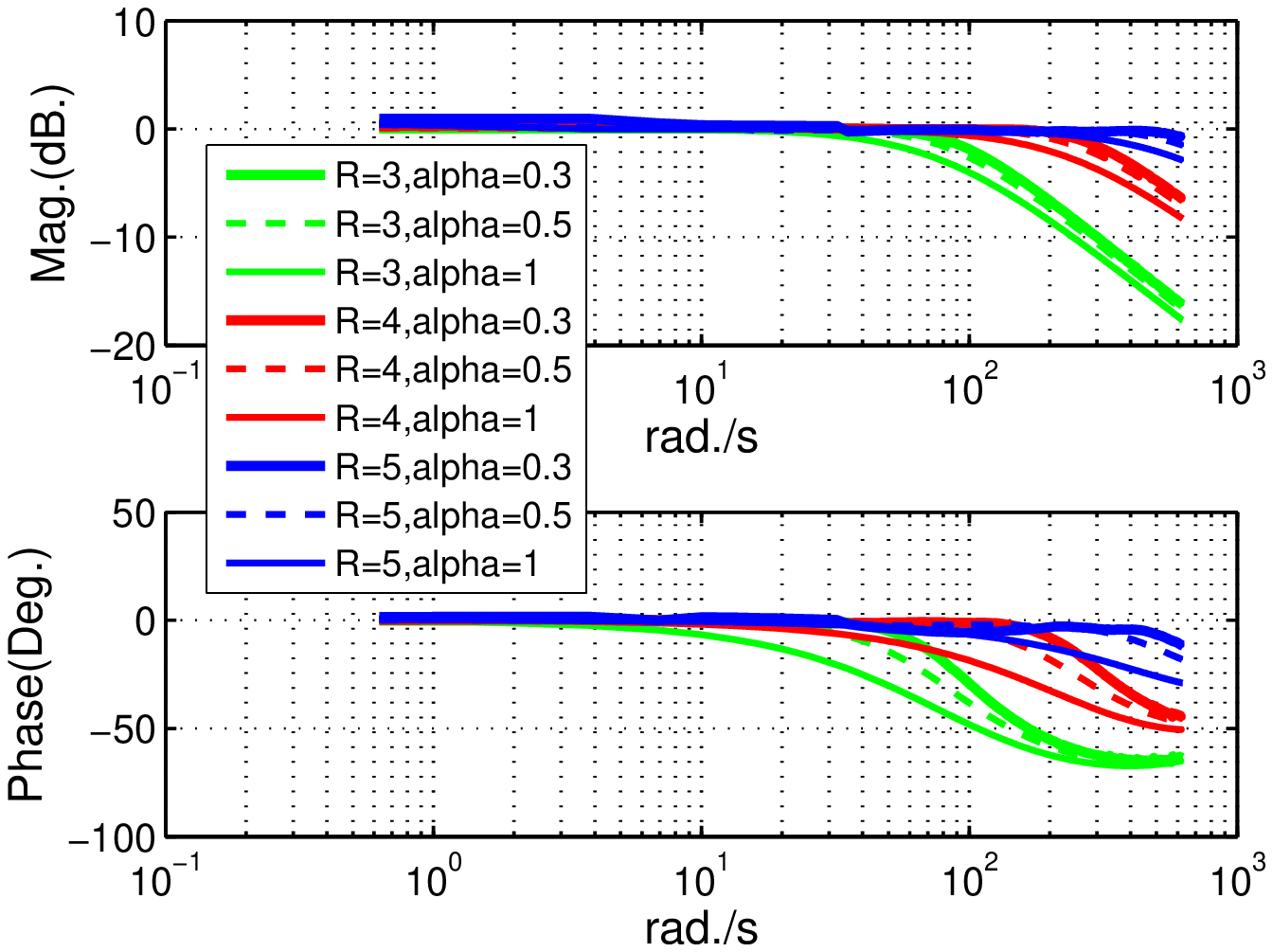}\\[0pt]
{\small 1(a)}\\[0pt]
\includegraphics[width=3.60in]{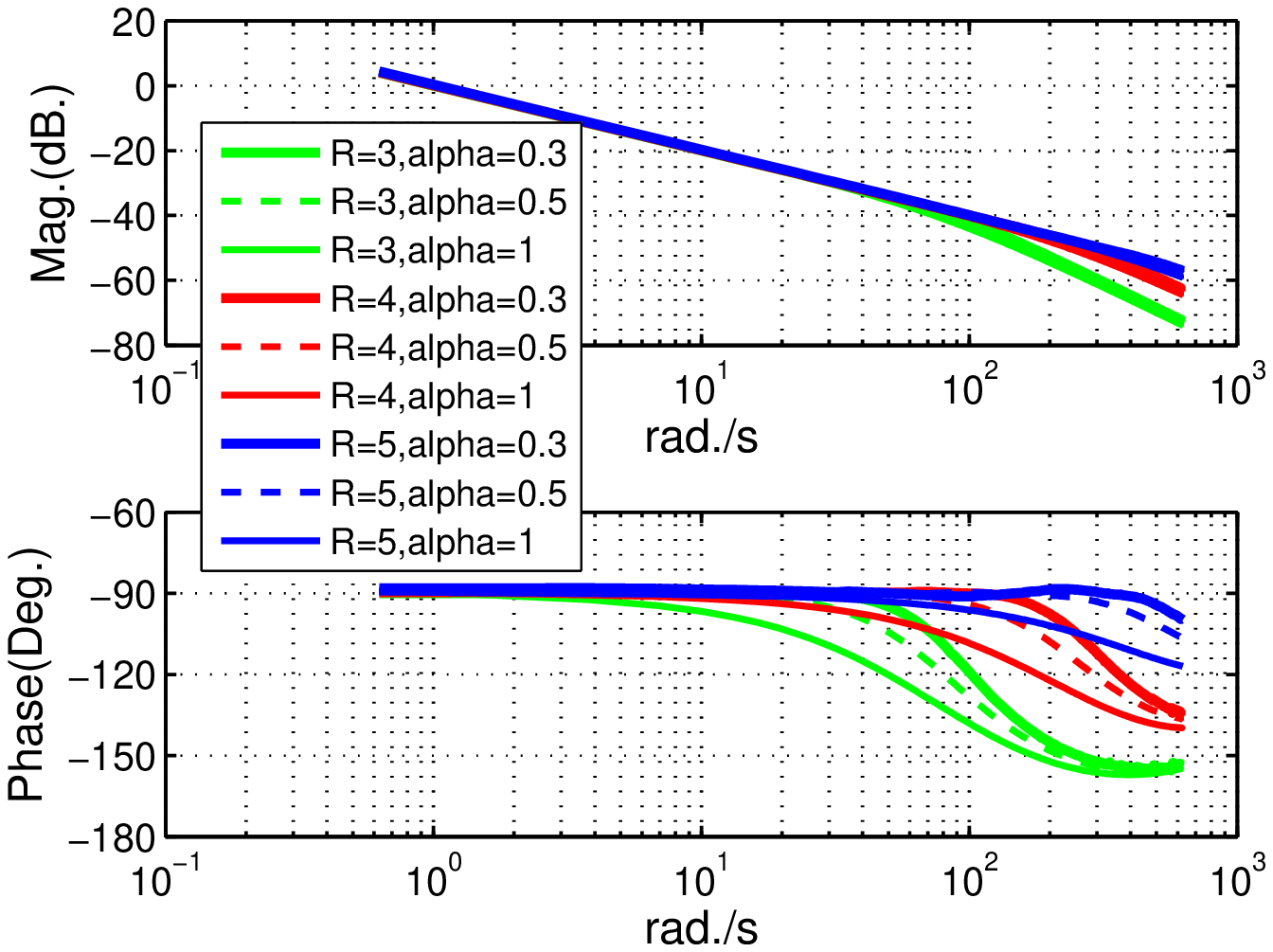}\\[0pt]
{\small 1(b)}\\[0pt]
\includegraphics[width=3.60in]{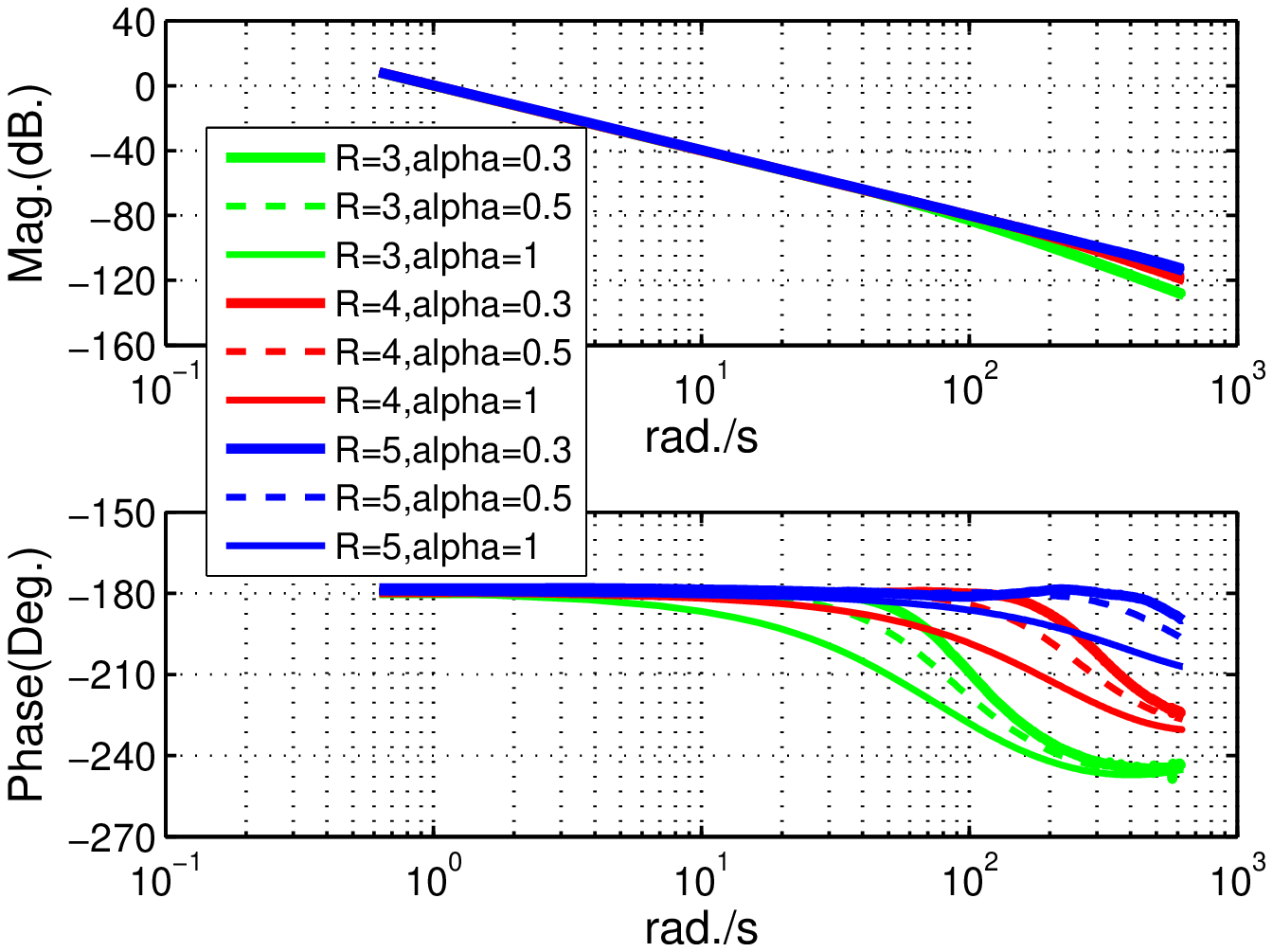}\\[0pt]
{\small 1(c)}
\end{center}
\caption{Frequency-domain characteristics with different $\protect%
\varepsilon $ and $\protect\alpha$. 1(a) Signal tracking. 1(b) Onefold
integral estimate. 1(c) Double integral estimate.}
\end{figure}

\begin{figure}[H]
\begin{center}
\includegraphics[width=3.60in]{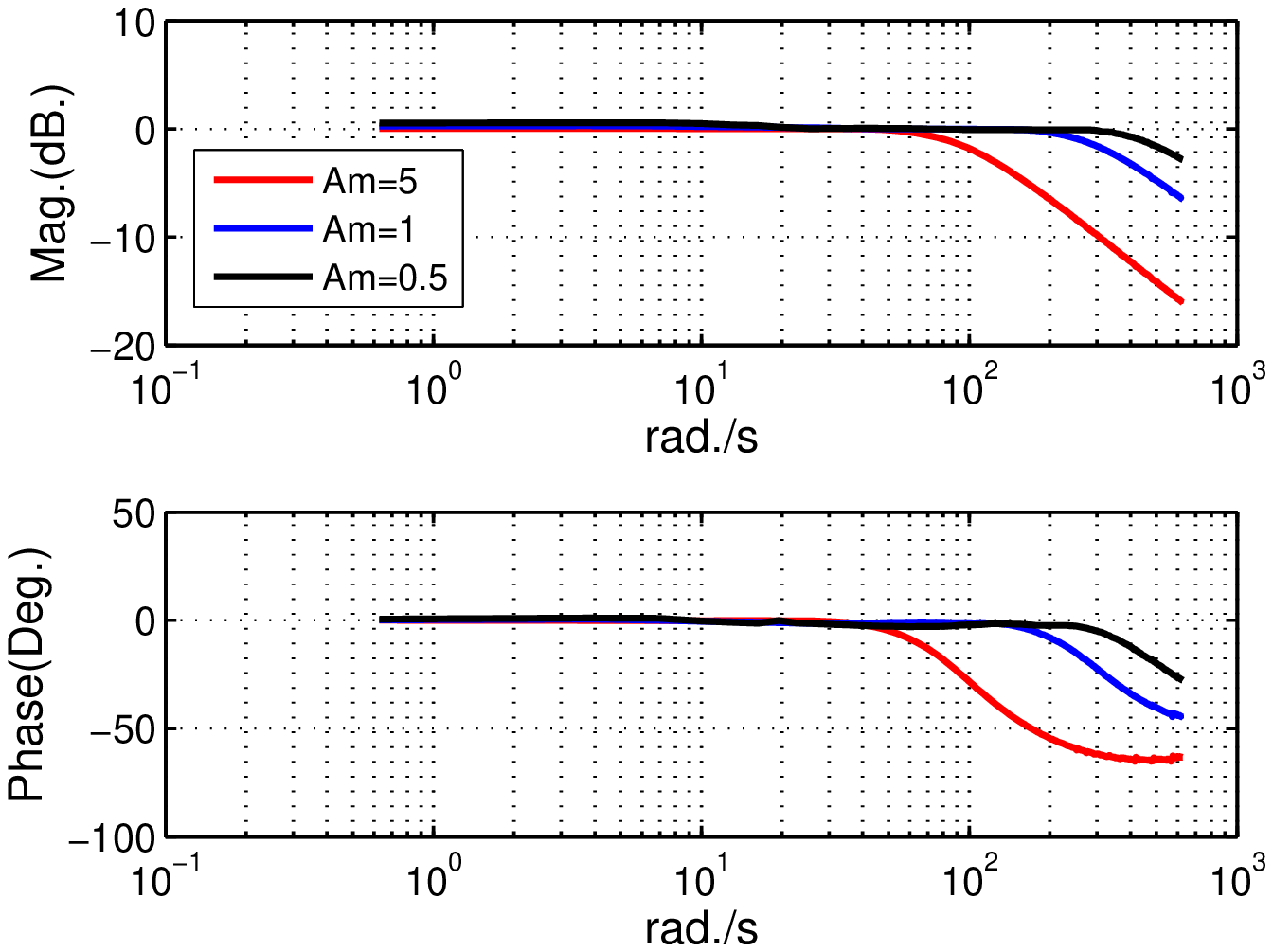}\\[0pt]
{\small 2(a)}\\[0pt]
\includegraphics[width=3.60in]{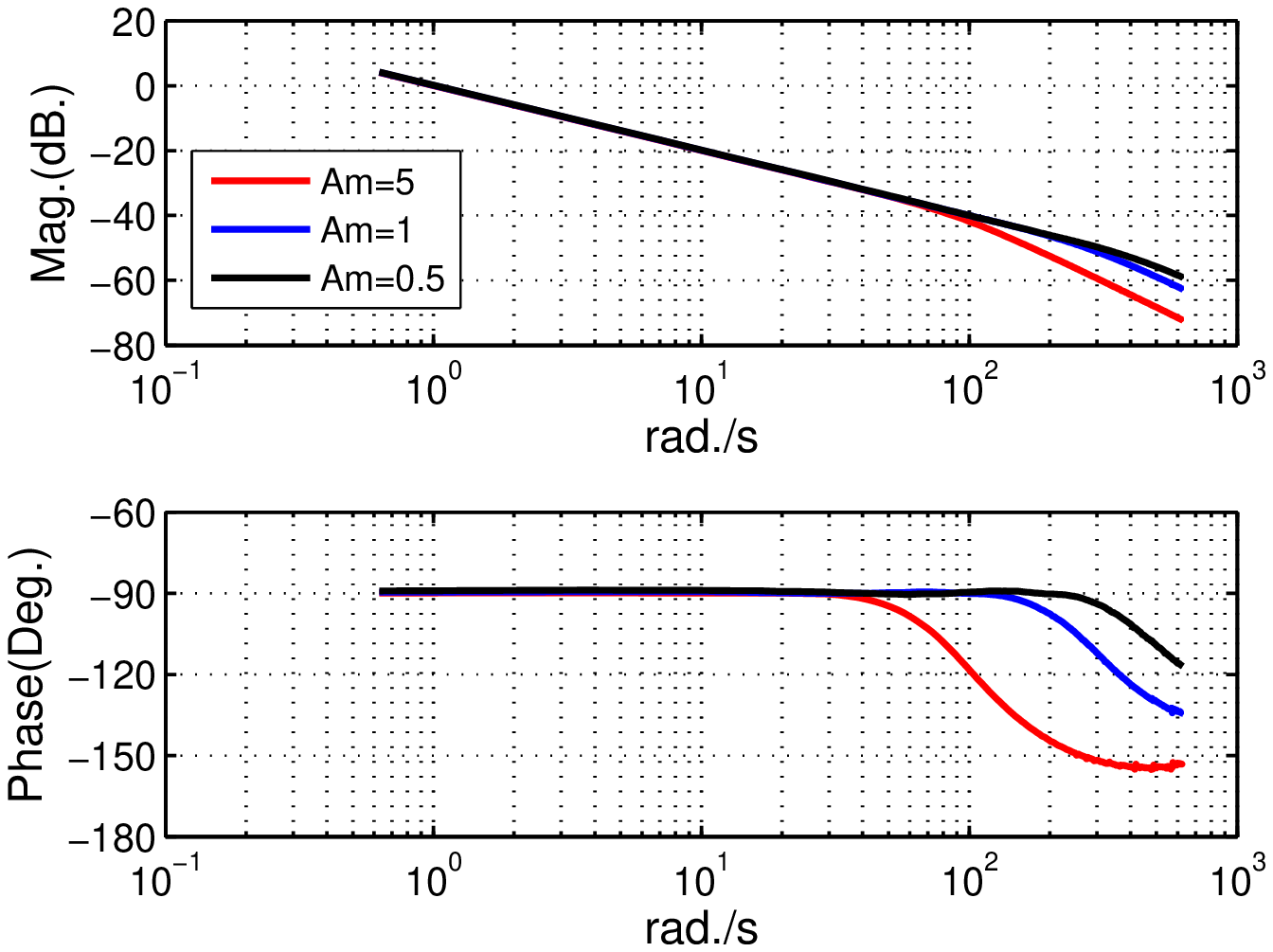}\\[0pt]
{\small 2(b)}\\[0pt]
\includegraphics[width=3.60in]{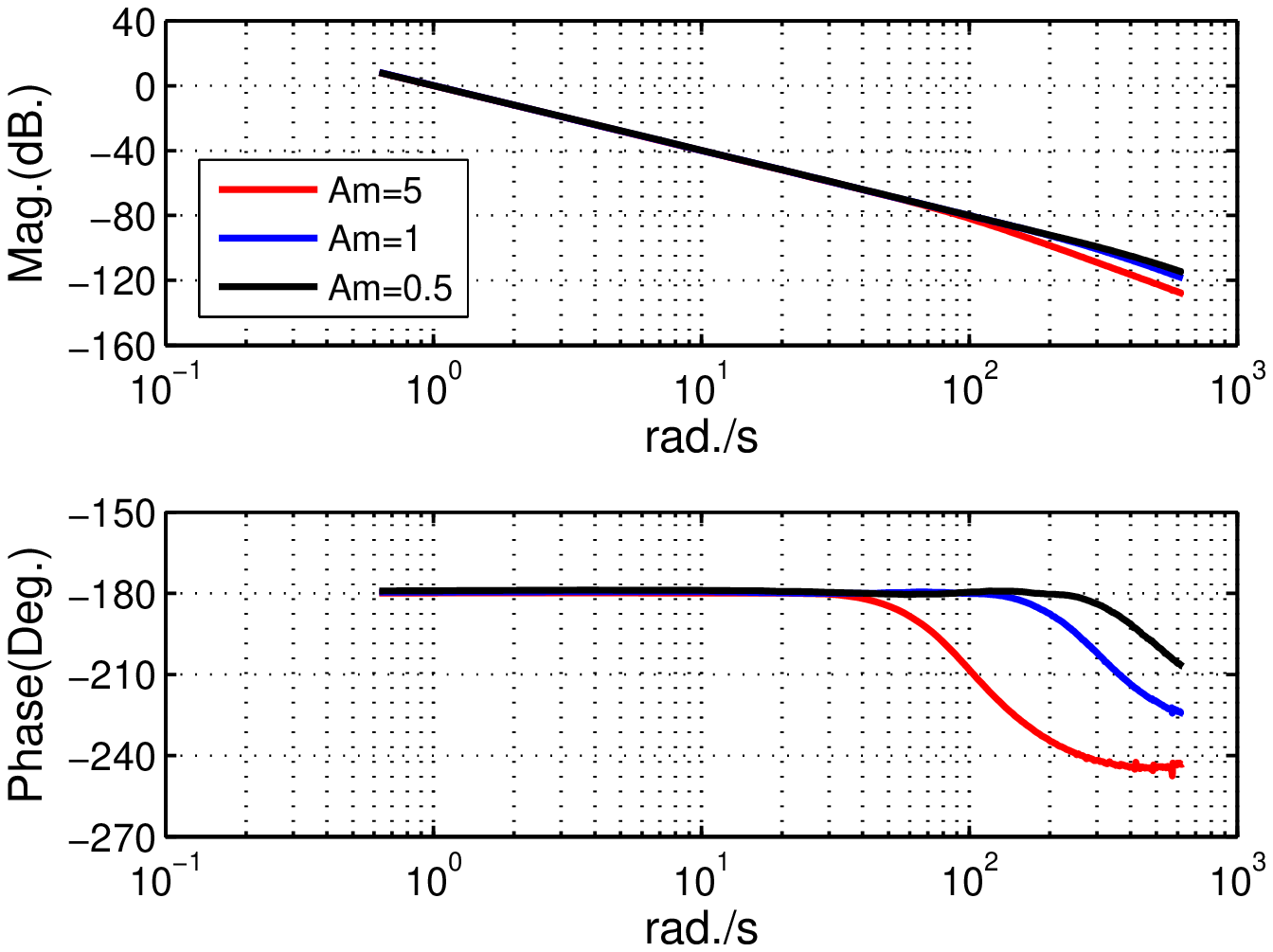}\\[0pt]
{\small 2(c)}
\end{center}
\caption{Frequency-domain characteristics with the change of $A_{m}$. 2(a)
Signal tracking. 2(b) Onefold integral estimate. 2(c) Double integral
estimate.}
\end{figure}

\begin{figure}[H]
\begin{center}
\includegraphics[width=2.80in]{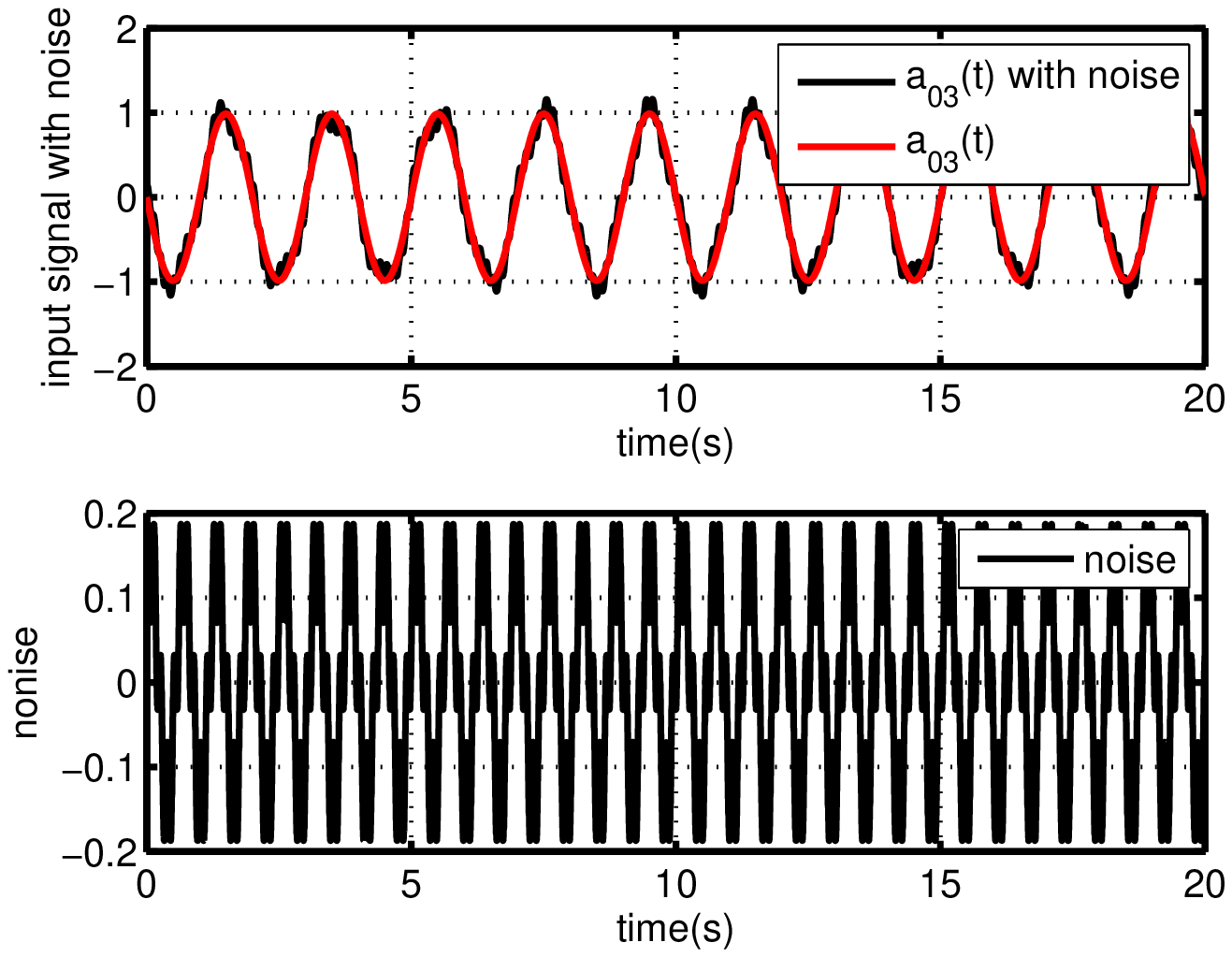}\\[0pt]
{\small 3(a)}\\[0pt]
\includegraphics[width=2.80in]{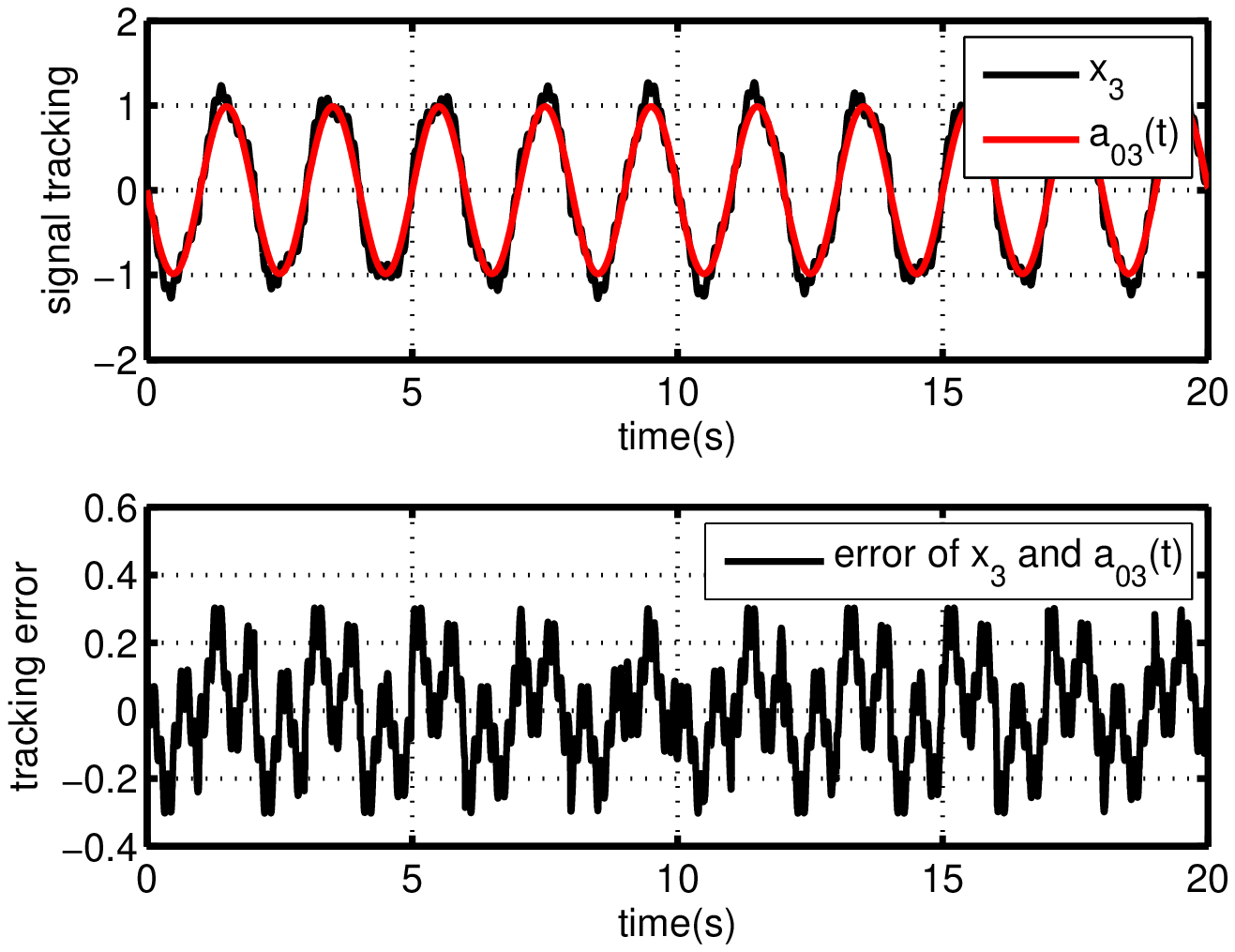}\\[0pt]
{\small 3(b)}\\[0pt]
\includegraphics[width=2.80in]{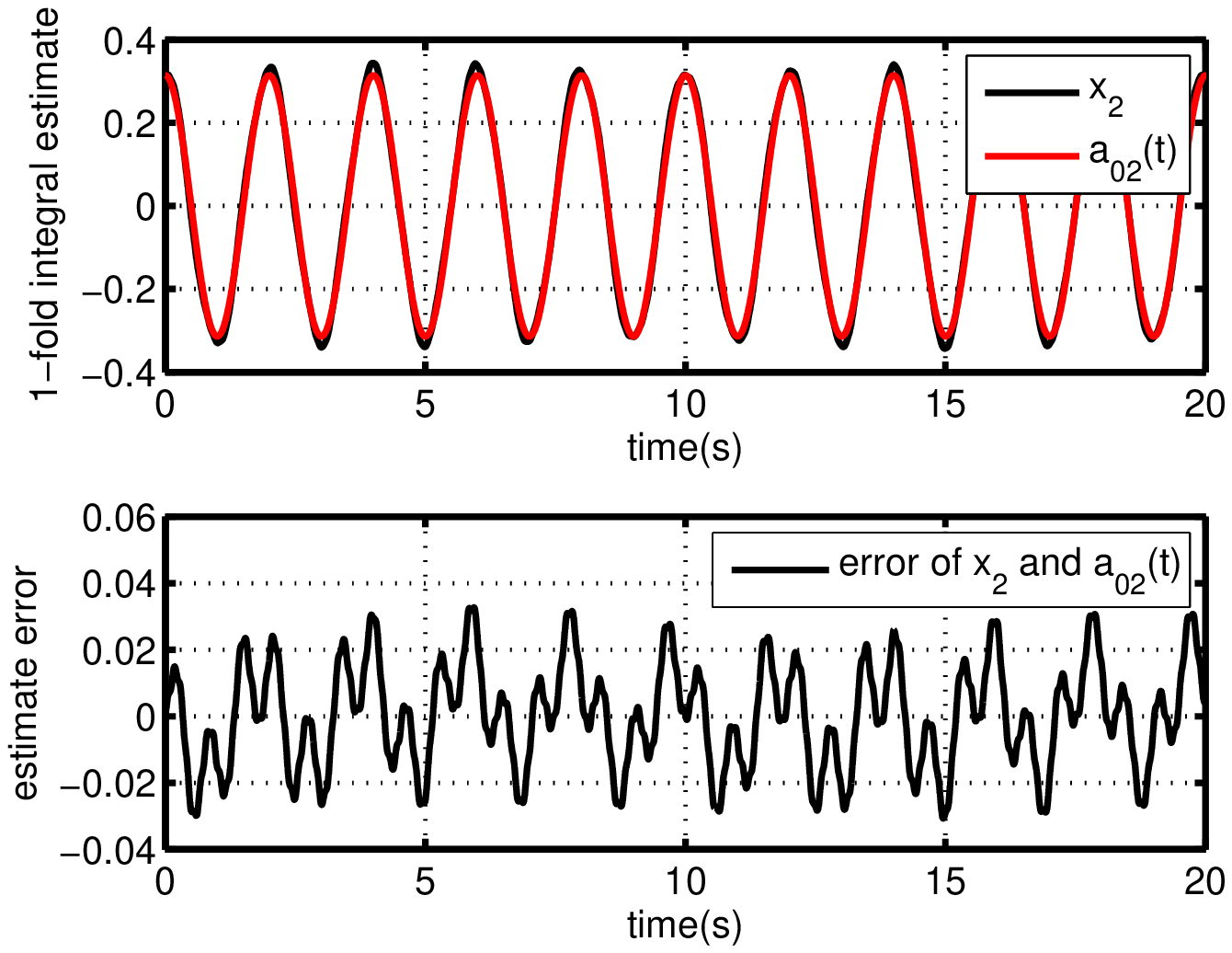}\\[0pt]
{\small 3(c)}\\[0pt]
\includegraphics[width=2.80in]{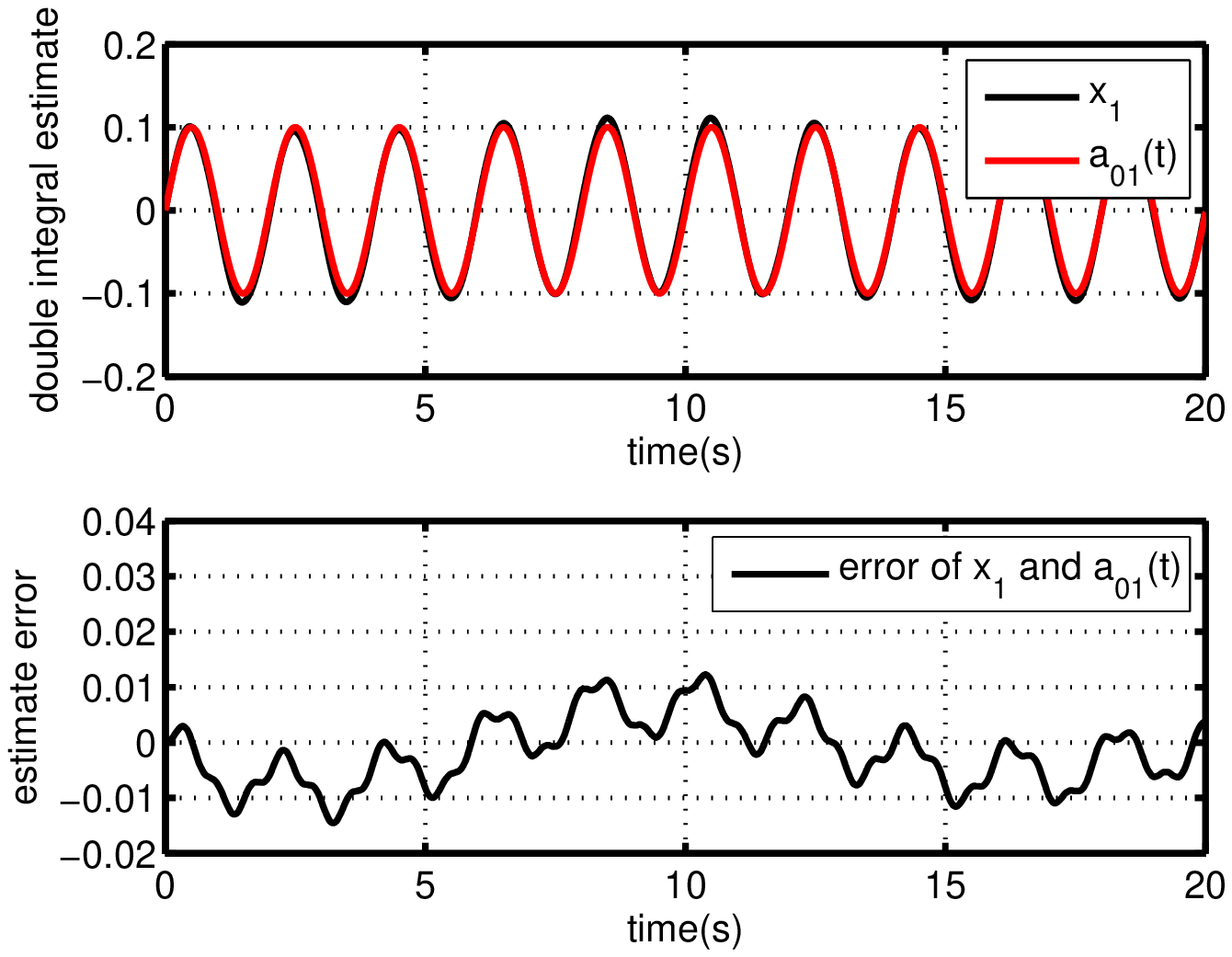}\\[0pt]
{\small 3(d)}
\end{center}
\caption{Integral estimations in 20s ($\protect\alpha _{3}=0.3$). 3(a) Input
signal with noise. 3(b) Signal tracking. 3(c) Onefold integral estimate. 3(d)
Double integral estimate.}
\end{figure}

\begin{figure}[H]
\begin{center}
\includegraphics[width=2.80in]{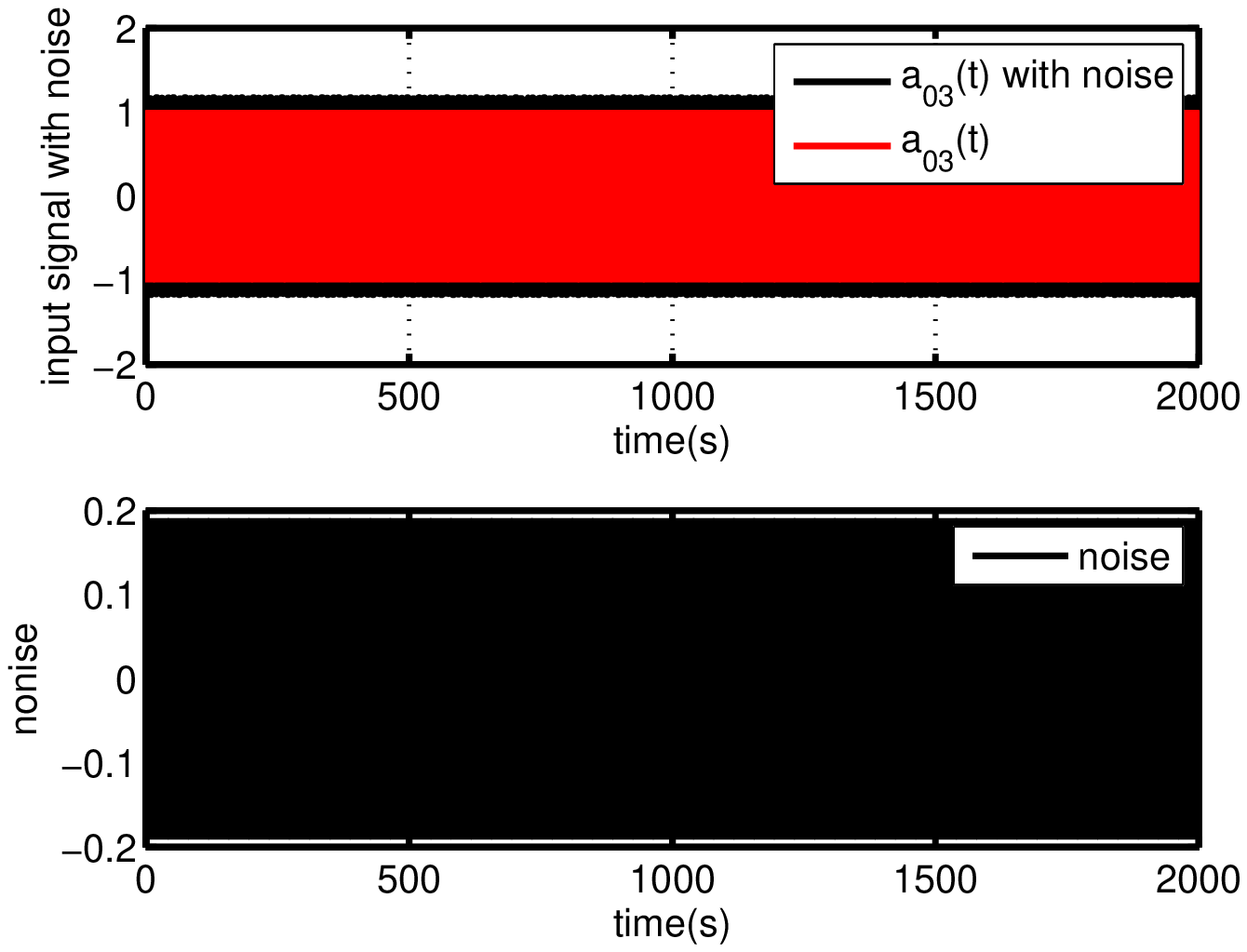}\\[0pt]
{\small 4(a)}\\[0pt]
\includegraphics[width=2.80in]{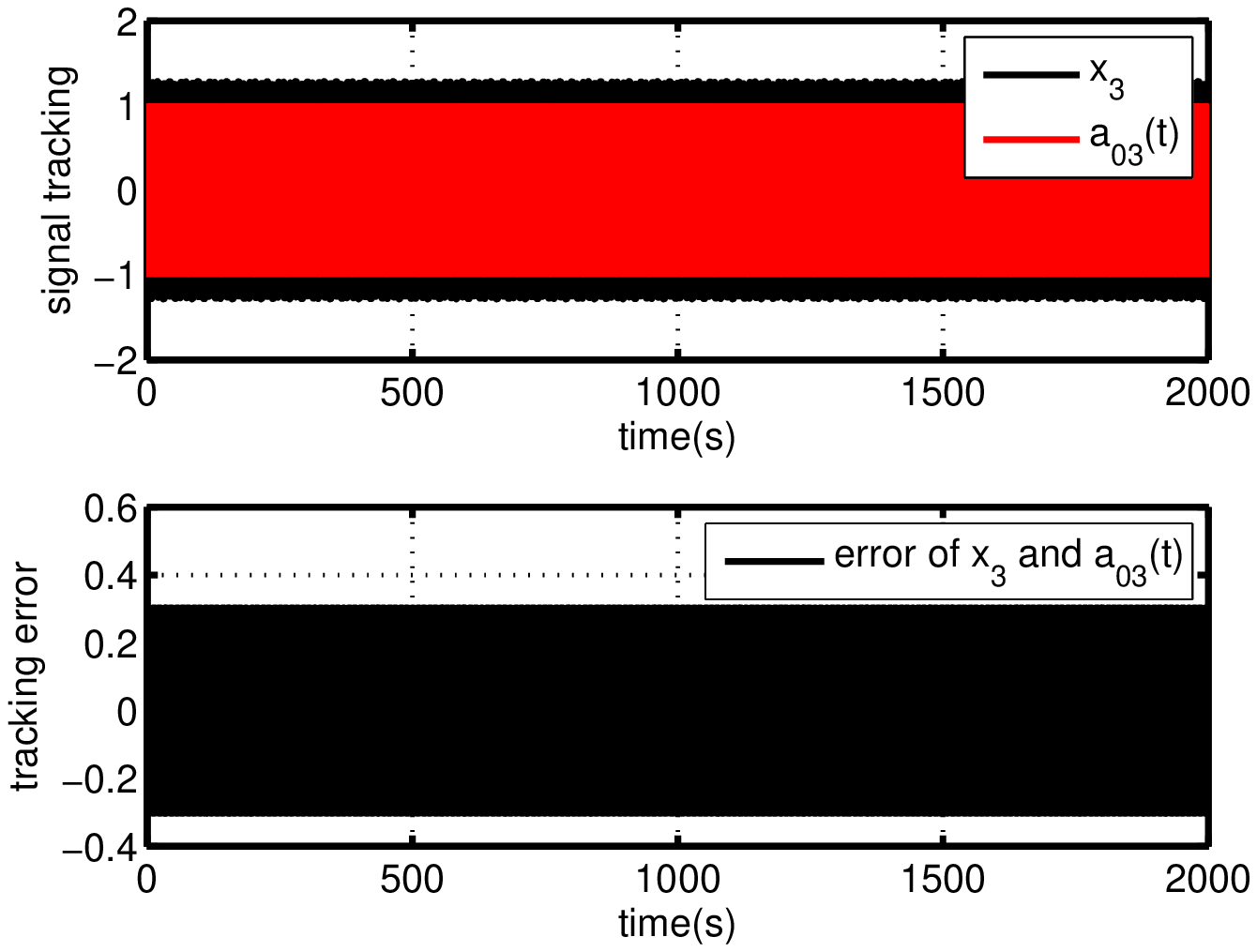}\\[0pt]
{\small 4(b)}\\[0pt]
\includegraphics[width=2.80in]{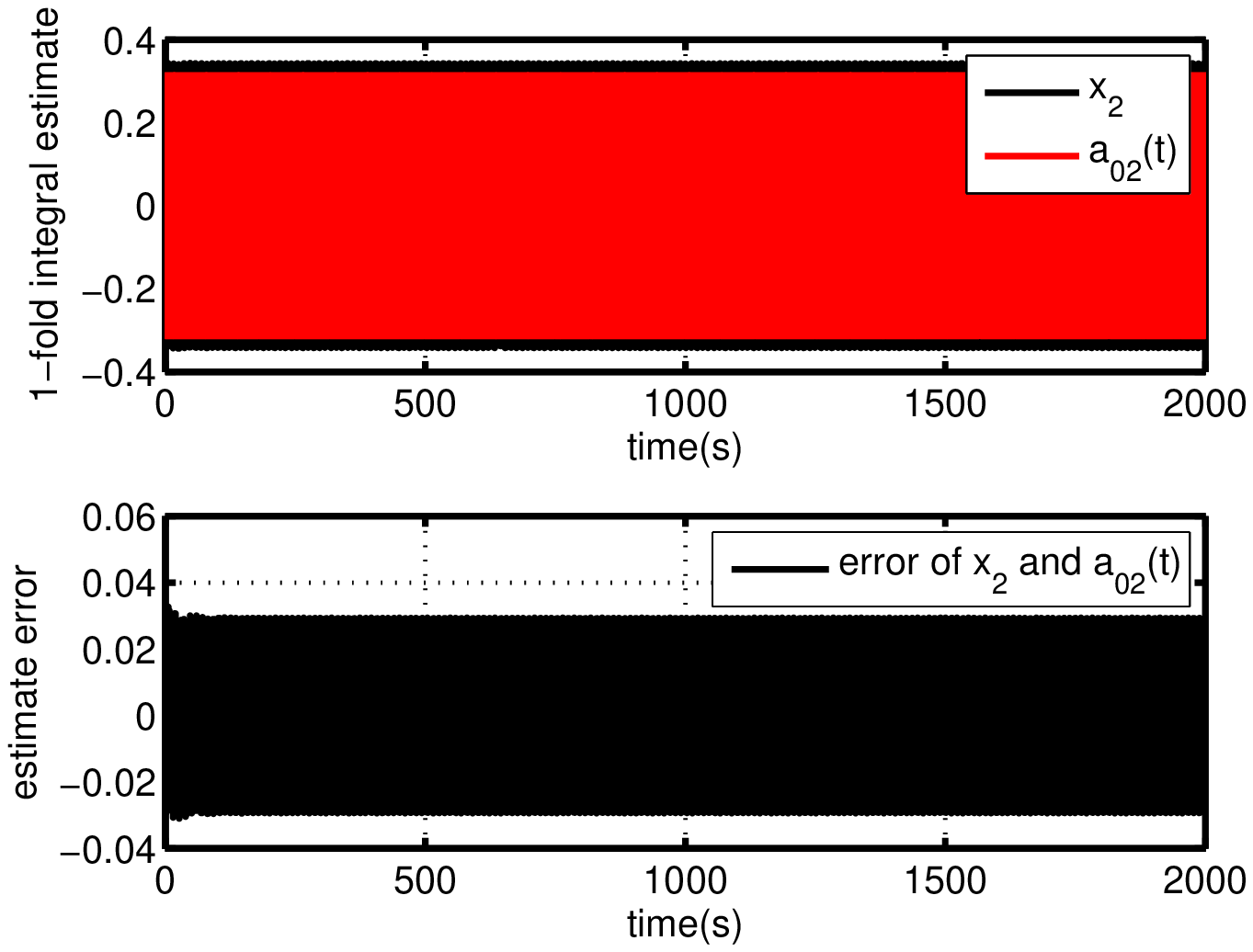}\\[0pt]
{\small 4(c)}\\[0pt]
\includegraphics[width=2.80in]{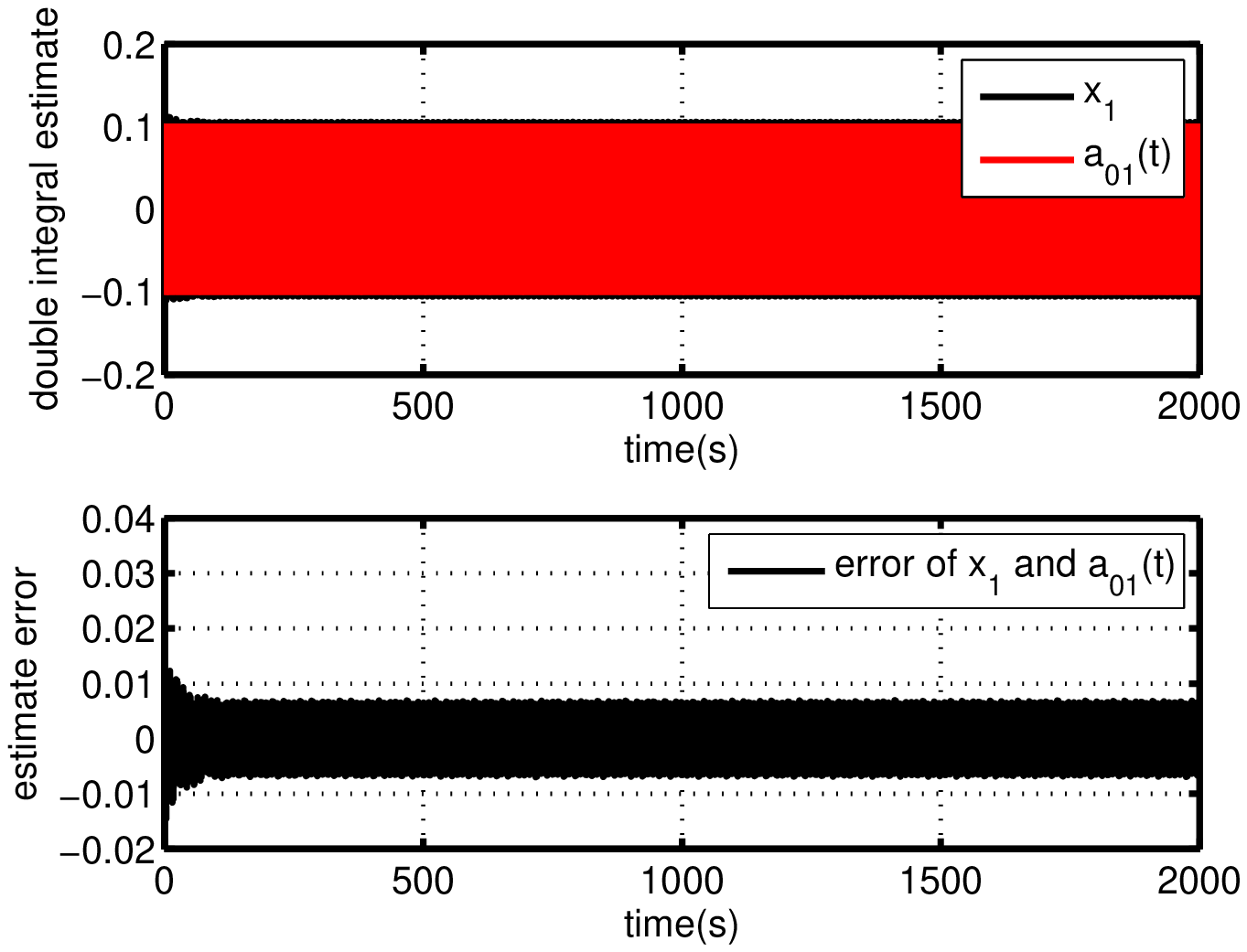}\\[0pt]
{\small 4(d)}
\end{center}
\caption{Integral estimations in 2000s ($\protect\alpha _{3}=0.3$). 4(a)
Input signal with noise. 4(b) Signal tracking. 4(c) Onefold integral
estimate. 4(d) Double integral estimate.}
\end{figure}

\begin{figure}[H]
\begin{center}
\includegraphics[width=2.80in]{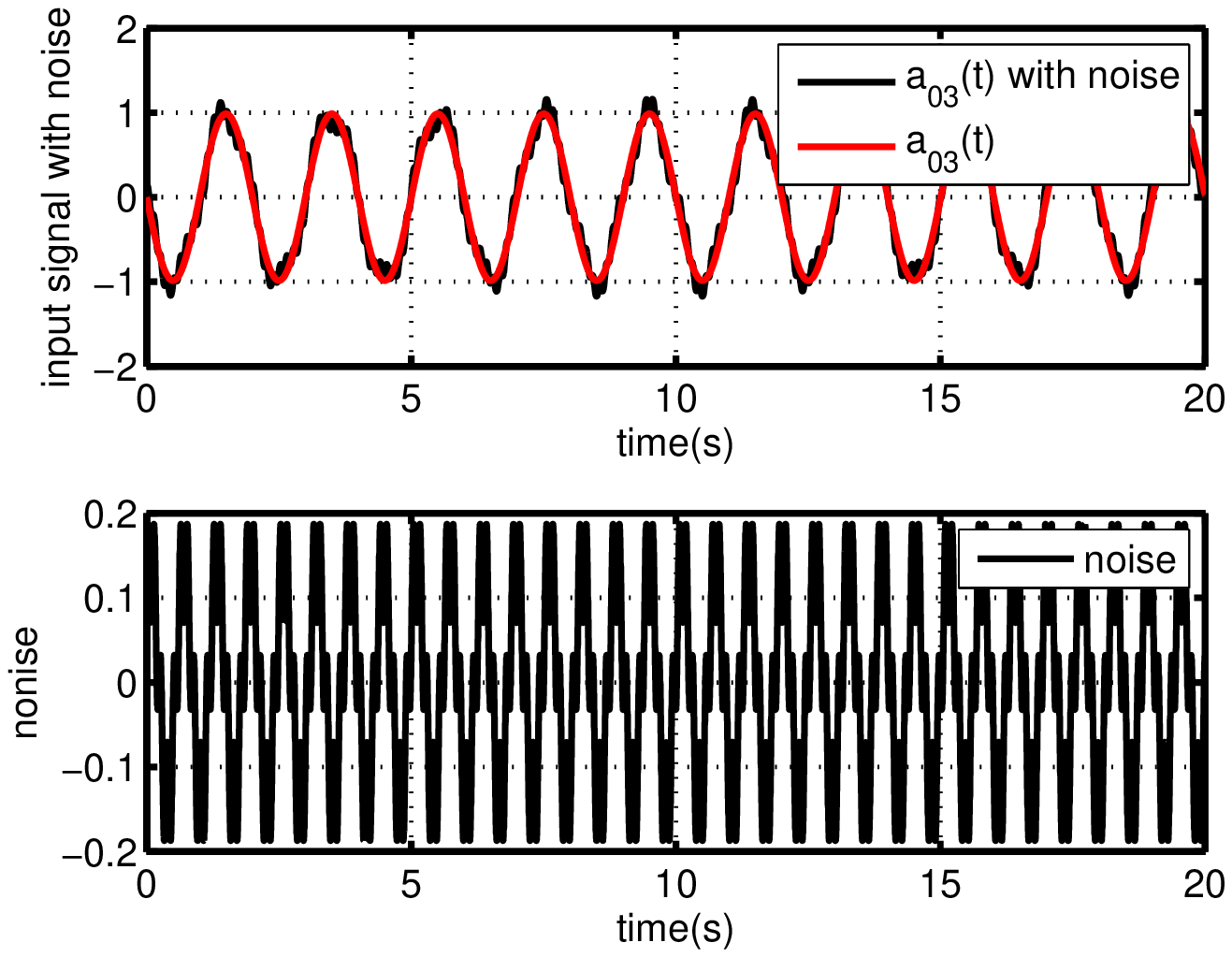}\\[0pt]
{\small 5(a)}\\[0pt]
\includegraphics[width=2.80in]{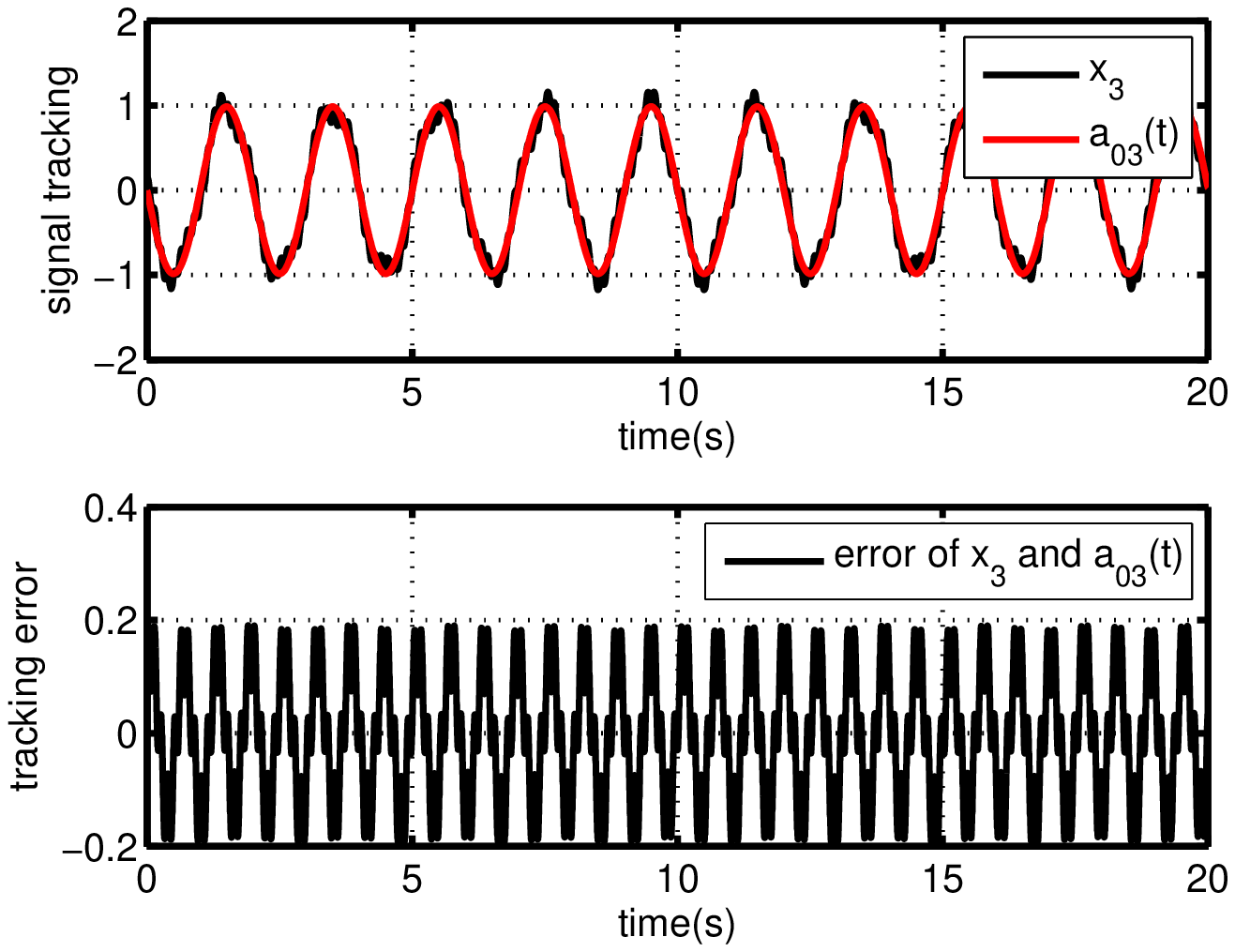}\\[0pt]
{\small 5(b)}\\[0pt]
\includegraphics[width=2.80in]{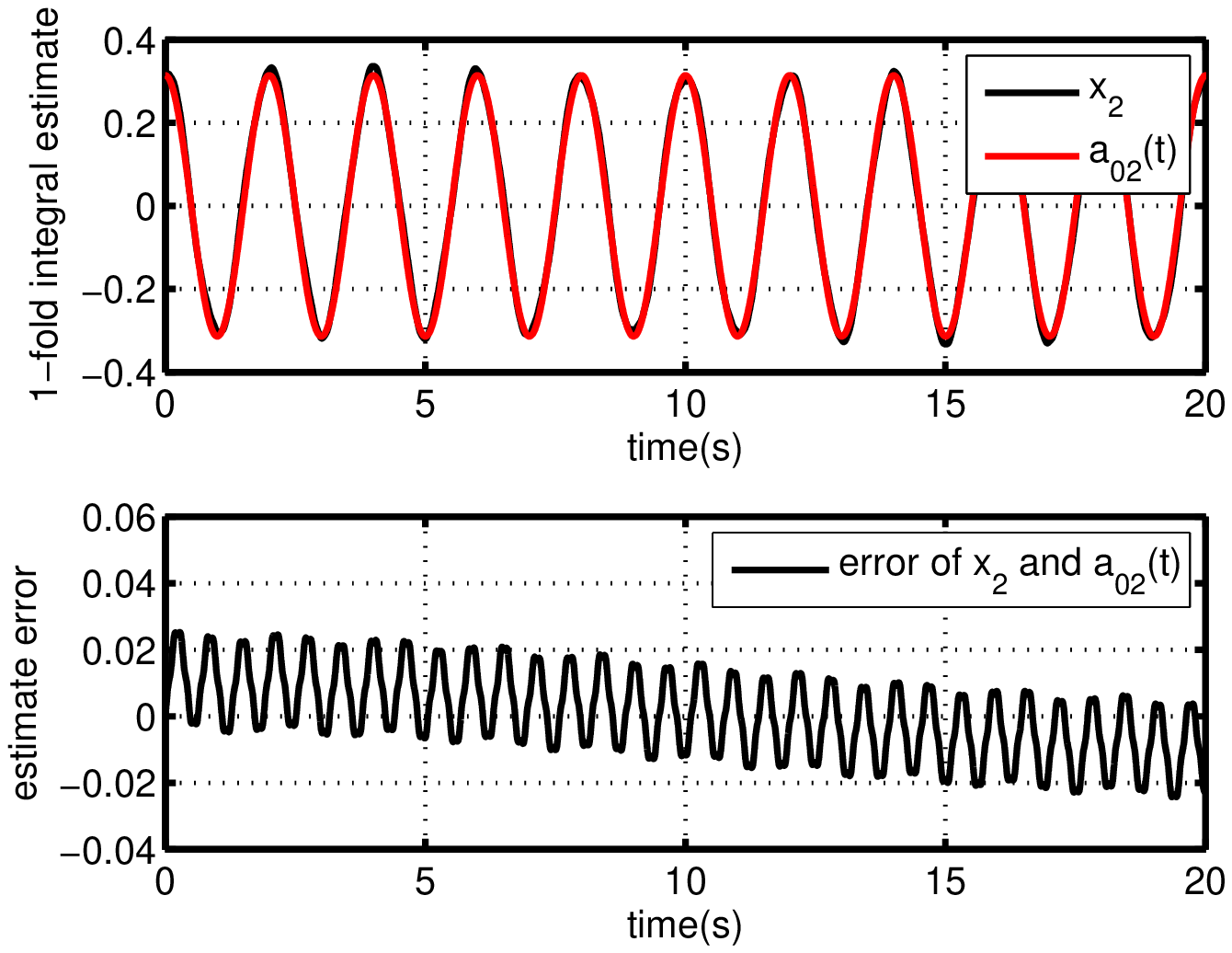}\\[0pt]
{\small 5(c)}\\[0pt]
\includegraphics[width=2.80in]{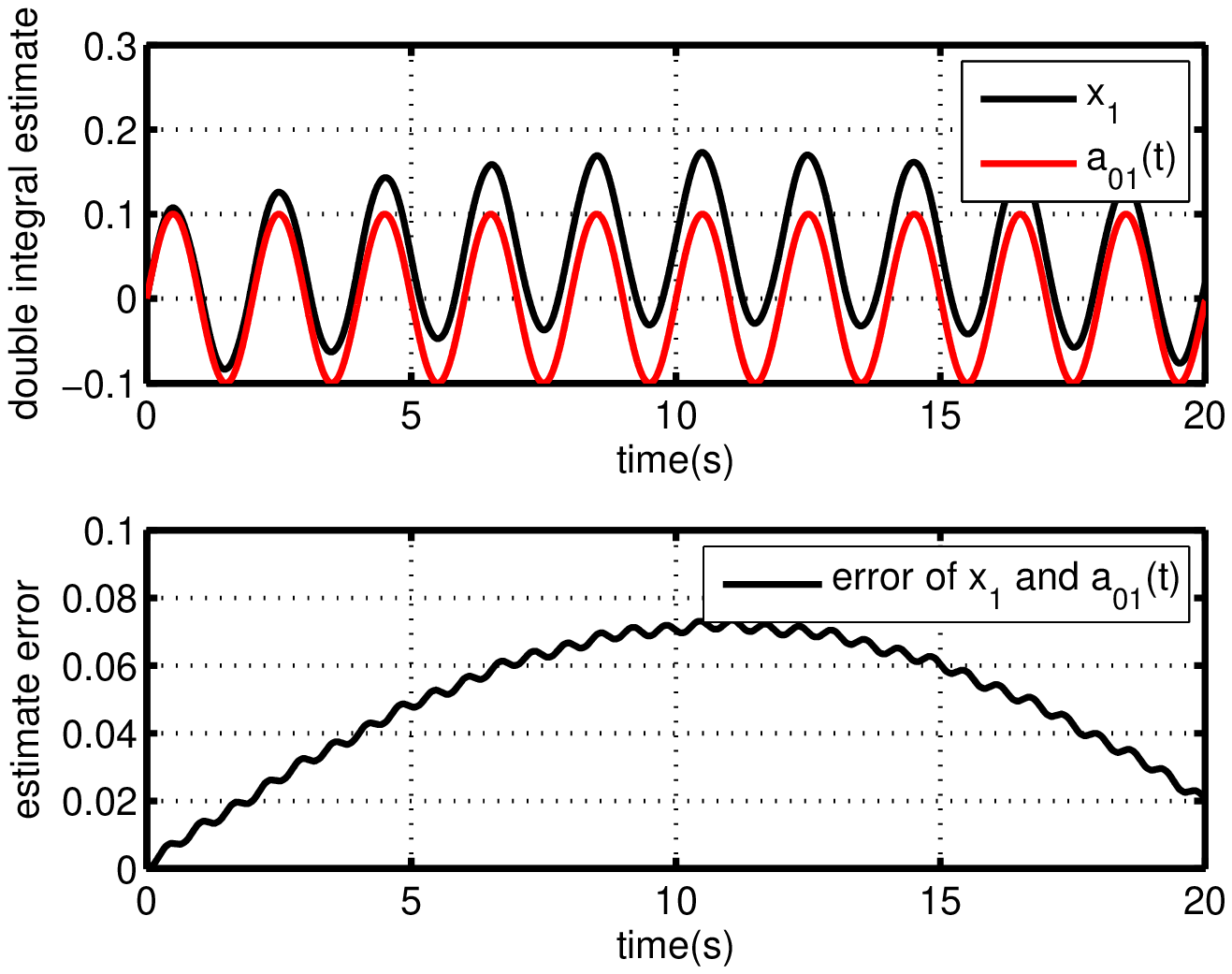}\\[0pt]
{\small 5(d)}
\end{center}
\caption{Integral estimations in 20s ($\protect\alpha _{3}=1$). 5(a) Input
signal with noise. 5(b) Signal tracking. 5(c) Onefold integral estimate. 5(d)
Double integral estimate.}
\end{figure}

\begin{figure}[H]
\begin{center}
\includegraphics[width=2.80in]{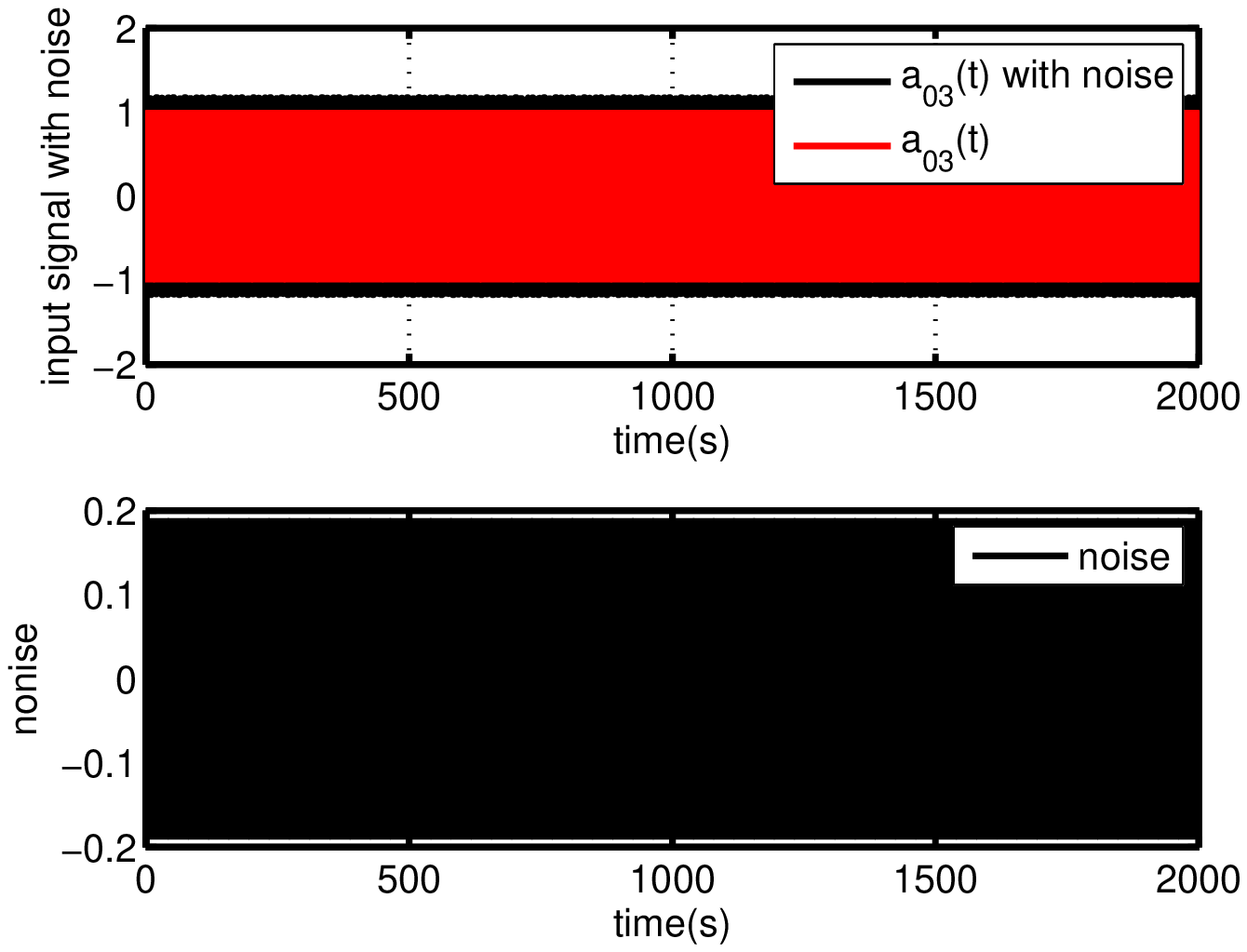}\\[0pt]
{\small 6(a)}\\[0pt]
\includegraphics[width=2.80in]{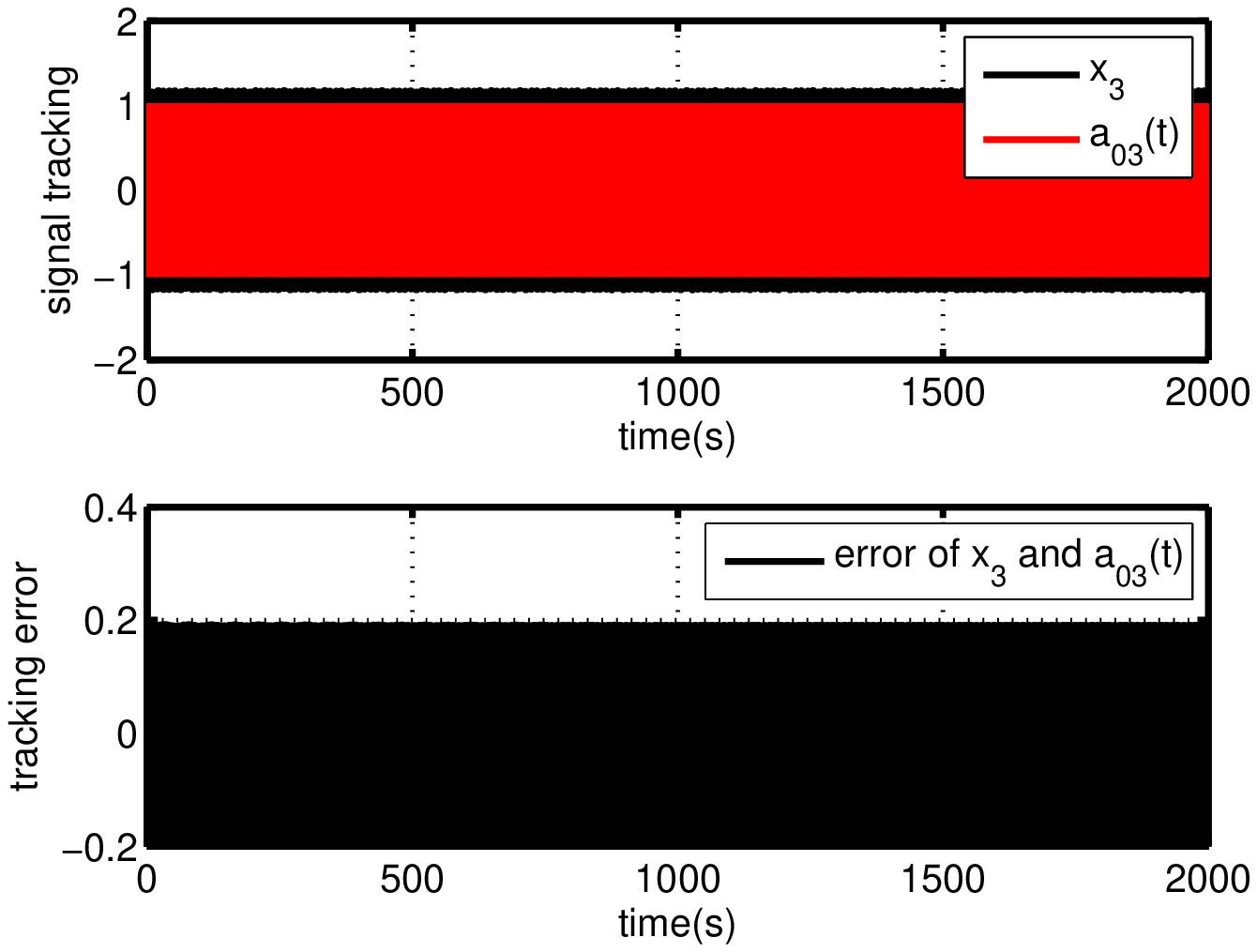}\\[0pt]
{\small 6(b)}\\[0pt]
\includegraphics[width=2.80in]{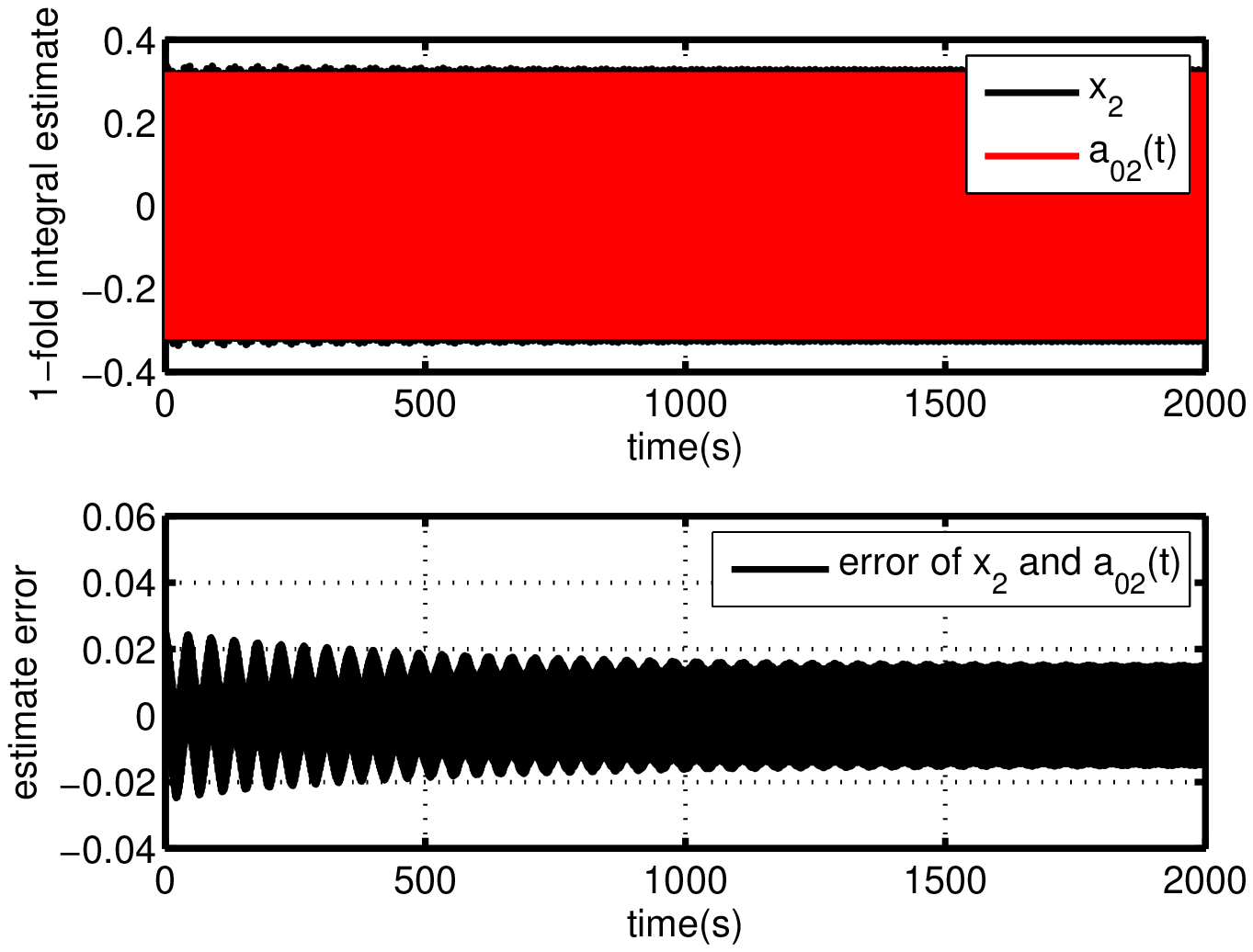}\\[0pt]
{\small 6(c)}\\[0pt]
\includegraphics[width=2.80in]{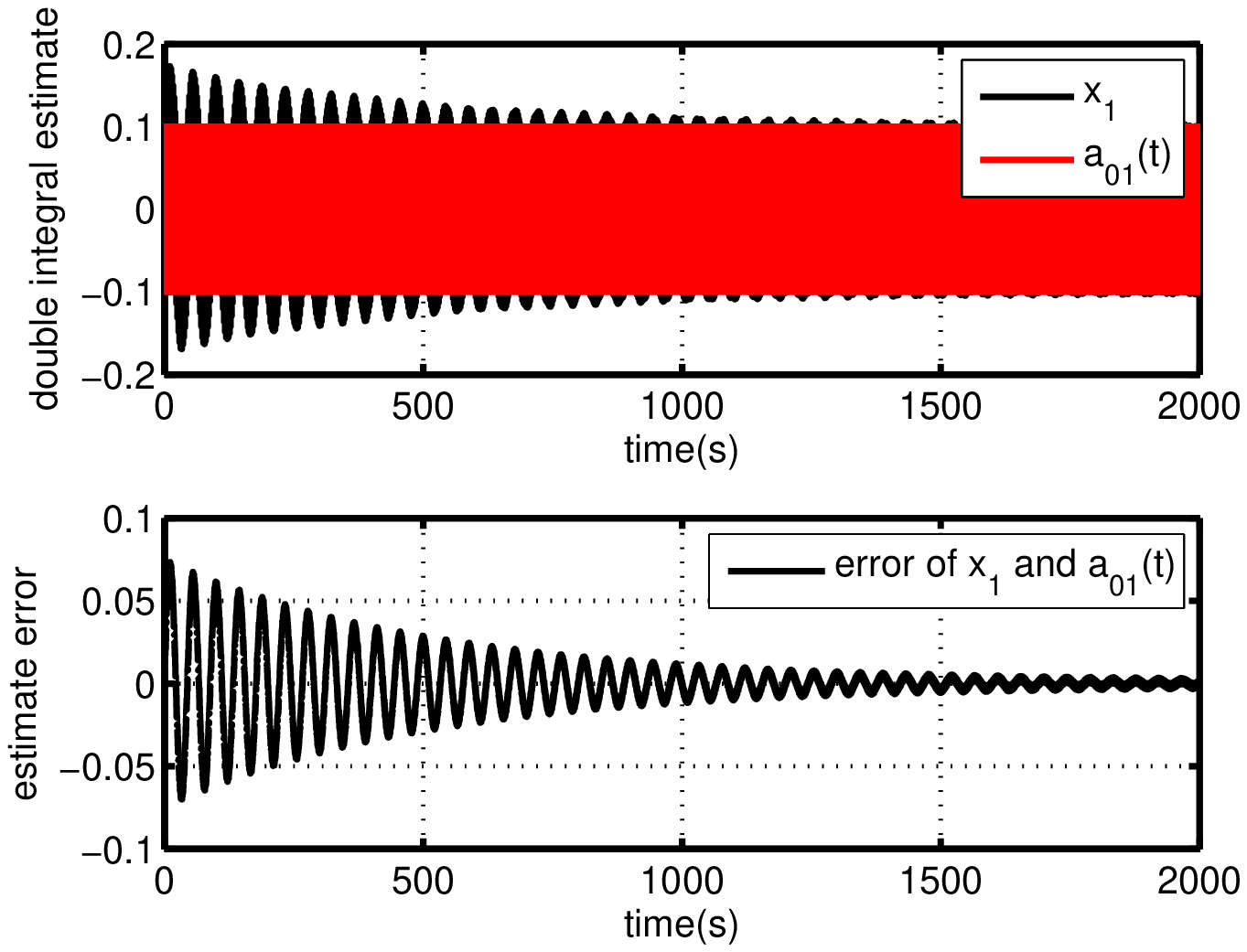}\\[0pt]
{\small 6(d)}
\end{center}
\caption{Integral estimations in 2000s ($\protect\alpha _{3}=1$). 6(a) Input
signal with noise. 6(b) Signal tracking. 6(c) Onefold integral estimate. 6(d)
Double integral estimate.}
\end{figure}

\end{document}